\documentclass[]{spie}  %>>> use for US letter paper
\usepackage{amsmath,amsfonts,amssymb, mathrsfs}
\usepackage{graphicx}
\usepackage[colorlinks=true, allcolors=blue]{hyperref}
\usepackage{makecell}
\usepackage{subfig}

\title{Characterization and performance of the second-year SPT-3G focal plane}

\author[a,b]{D.~Dutcher}
\author[c]{P.~A.~R.~Ade}
\author[d,e]{Z.~Ahmed}
\author[f,a]{A.~J.~Anderson}
\author[g]{J.~S.~Avva}
\author[a]{R.~Basu Thakur}
\author[h,a]{A.~N.~Bender}
\author[f,a,i]{B.~A.~Benson}
\author[a,j,b,h,i]{J.~E.~Carlstrom}
\author[h,a]{F.~W.~Carter}
\author[h]{T.~W.~Cecil}
\author[h,a,i]{C.~L.~Chang}
\author[k]{J.~F.~Cliche}
\author[g]{A.~Cukierman}
\author[g]{T.~de~Haan}
\author[l]{J.~Ding}
\author[k,m]{M.~A.~Dobbs}
\author[n]{W.~Everett}
\author[o]{A.~Foster}
\author[a,p]{J.~Gallicchio}
\author[k]{A.~Gilbert}
\author[g]{J.~C.~Groh}
\author[g]{S.~T.~Guns}
\author[n,q]{N.~W.~Halverson}
\author[h,r]{A.~H.~Harke-Hosemann}
\author[g]{N.~L.~Harrington}
\author[h,a]{J.~W.~Henning}
\author[g]{W.~L.~Holzapfel}
\author[g]{N.~Huang}
\author[d,s,e]{K.~D.~Irwin}
\author[g]{O.~B.~Jeong}
\author[f]{M.~Jonas}
\author[l]{T.~S.~Khaire}
\author[r]{A.~M.~Kofman}
\author[o]{M.~Korman}
\author[f]{D.~L.~Kubik}
\author[h]{S.~Kuhlmann}
\author[d,s,e]{C.-L.~Kuo}
\author[g,t]{A.~T.~Lee}
\author[a]{A.~E.~Lowitz}
\author[a,j,b,i]{S.~S.~Meyer}
\author[u]{D.~Michalik}
\author[k]{J.~Montgomery}
\author[r]{A.~Nadolski}
\author[v]{T.~Natoli}
\author[f]{H.~Nguyen}
\author[k]{G.~I.~Noble}
\author[l]{V.~Novosad}
\author[a]{S.~Padin}
\author[a,b]{Z.~Pan}
\author[l]{J.~Pearson}
\author[l]{C.~M.~Posada}
\author[a,b]{W.~Quan}
\author[f,a]{A.~Rahlin}
\author[o]{J.~E.~Ruhl}
\author[n]{J.T.~Sayre}
\author[a,i]{E.~Shirokoff}
\author[w]{G.~Smecher}
\author[a,b]{J.~A.~Sobrin}
\author[x]{A.~A.~Stark}
\author[d,s]{K.~T.~Story}
\author[t]{A.~Suzuki}
\author[d,s,e]{K.~L.~Thompson}
\author[c]{C.~Tucker}
\author[v,y]{K.~Vanderlinde}
\author[r,z]{J.~D.~Vieira}
\author[h]{G.~Wang}
\author[aa,g]{N.~Whitehorn}
\author[h]{V.~Yefremenko}
\author[d,s,e]{K.~W.~Yoon}
\author[y]{M.~R.~Young}

\affil[a]{Kavli Institute for Cosmological Physics, University of Chicago, 5640 South Ellis Avenue, Chicago, IL, USA 60637}
\affil[b]{Department of Physics, University of Chicago, 5640 South Ellis Avenue, Chicago, IL, USA 60637}
\affil[c]{School of Physics and Astronomy, Cardiff University, Cardiff CF24 3YB, United Kingdom}
\affil[d]{Kavli Institute for Particle Astrophysics and Cosmology, Stanford University, 452 Lomita Mall, Stanford, CA, USA 94305}
\affil[e]{SLAC National Accelerator Laboratory, 2575 Sand Hill Road, Menlo Park, CA, USA 94025}
\affil[f]{Fermi National Accelerator Laboratory, MS209, P.O. Box 500, Batavia, IL, USA 60510}
\affil[g]{Department of Physics, University of California, Berkeley, CA, USA 94720}
\affil[h]{High-Energy Physics Division, Argonne National Laboratory, 9700 South Cass Avenue., Argonne, IL, USA 60439}
\affil[i]{Department of Astronomy and Astrophysics, University of Chicago, 5640 South Ellis Avenue, Chicago, IL, USA 60637}
\affil[j]{Enrico Fermi Institute, University of Chicago, 5640 South Ellis Avenue, Chicago, IL, USA 60637}
\affil[k]{Department of Physics and McGill Space Institute, McGill University, 3600 Rue University, Montreal, Quebec H3A 2T8, Canada}
\affil[l]{Materials Sciences Division, Argonne National Laboratory, 9700 South Cass Avenue, Argonne, IL, USA 60439}
\affil[m]{Canadian Institute for Advanced Research, CIFAR Program in Gravity and the Extreme Universe, Toronto, ON, M5G 1Z8, Canada}
\affil[n]{CASA, Department of Astrophysical and Planetary Sciences, University of Colorado, Boulder, CO, USA 80309}
\affil[o]{Department of Physics, Center for Education and Research in Cosmology and Astrophysics, Case Western Reserve University, Cleveland, OH, USA 44106}
\affil[p]{Harvey Mudd College, 301 Platt Boulevard., Claremont, CA, USA 91711}
\affil[q]{Department of Physics, University of Colorado, Boulder, CO, USA 80309}
\affil[r]{Department of Astronomy, University of Illinois at Urbana-Champaign, 1002 West Green Street, Urbana, IL, USA 61801}
\affil[s]{Deptartment of Physics, Stanford University, 382 Via Pueblo Mall, Stanford, CA, USA 94305}
\affil[t]{Physics Division, Lawrence Berkeley National Laboratory, Berkeley, CA, USA 94720}
\affil[u]{University of Chicago, 5640 South Ellis Avenue, Chicago, IL, USA 60637}
\affil[v]{Dunlap Institute for Astronomy \& Astrophysics, University of Toronto, 50 St. George Street, Toronto, ON, M5S 3H4, Canada}
\affil[w]{Three-Speed Logic, Inc., Vancouver, B.C., V6A 2J8, Canada}
\affil[x]{Harvard-Smithsonian Center for Astrophysics, 60 Garden Street, Cambridge, MA, USA 02138}
\affil[y]{Department of Astronomy \& Astrophysics, University of Toronto, 50 St. George Street, Toronto, ON, M5S 3H4, Canada}
\affil[z]{Department of Physics, University of Illinois Urbana-Champaign, 1110 West Green Street, Urbana, IL, USA 61801}
\affil[aa]{Department of Physics and Astronomy, University of California, Los Angeles, CA, USA 90095}

\authorinfo{Send correspondence to D. Dutcher: E-mail: ddutcher@uchicago.edu}

\begin{document} 
\maketitle

\begin{abstract}
The third-generation instrument for the 10-meter South Pole Telescope, SPT-3G, was first installed in January 2017. In addition to completely new cryostats, secondary telescope optics, and readout electronics, the number of detectors in the focal plane has increased by an order of magnitude from previous instruments to $\sim$16,000.  The SPT-3G focal plane consists of ten detector modules, each with an array of 269 trichroic, polarization-sensitive pixels on a six-inch silicon wafer. Within each pixel is a broadband, dual-polarization sinuous antenna; the signal from each orthogonal linear polarization is divided into three frequency bands centered at 95, 150, and 220~GHz by in-line lumped element filters and transmitted via superconducting microstrip to Ti/Au transition-edge sensor (TES) bolometers. Properties of the TES film, microstrip filters, and bolometer island must be tightly controlled to achieve optimal performance. For the second year of SPT-3G operation, we have replaced all ten wafers in the focal plane with new detector arrays tuned to increase mapping speed and improve overall performance.  Here we discuss the TES superconducting transition temperature and normal resistance, detector saturation power, bandpasses, optical efficiency, and full array yield for the 2018 focal plane.
\end{abstract}

\keywords{Cosmology, CMB, TES, detectors, SPT}

\section{INTRODUCTION}

SPT-3G is the third survey instrument to be deployed on the South Pole Telescope, a 10-meter mm-wavelength telescope located at the Amundsen-Scott South Pole Station and designed for high angular-resolution observations of the cosmic microwave background\cite{Carlstrom}. A major upgrade over its predecessors, SPT-3G includes changes to several aspects of the telescope:\cite{Benson}$^,$\cite{Anderson} the secondary telescope optics were redesigned to accommodate a wide-aperture coupling that maximizes the illuminated focal plane area, new optics and receiver cryostats were built to house the larger lenses and detector cold stage\cite{Joshua}, and the readout electronics were replaced with a new system with a 4x-higher multiplexing factor\cite{Bender14}. To take full advantage of the increased focal plane area and readout capability, a new generation of detectors populate the focal plane.

\begin{figure}
\begin{center}
\subfloat{{\includegraphics[height=3.8cm]{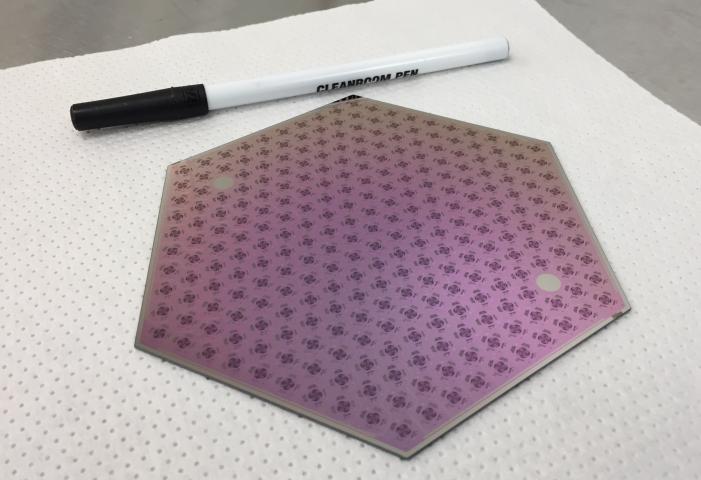} }}
\qquad
\subfloat{{\includegraphics[height=3.8cm]{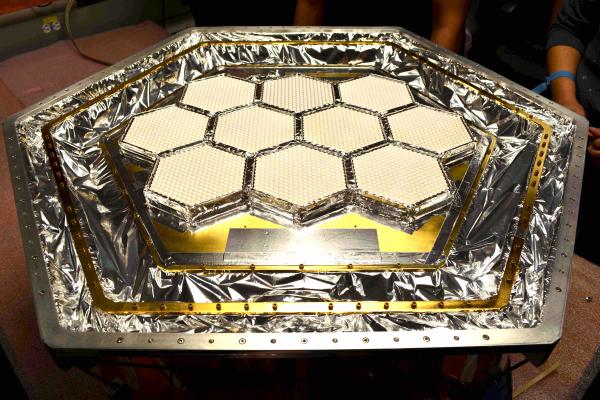} }}
\end{center}
\caption[focal_plane]
{\label{fig:focal_plane}
(Left) An SPT-3G detector array. The antenna of each pixel is visible. (Right) The SPT-3G focal plane, containing ten modules of detector and lenslet\cite{Andrew} arrays and their associated readout installed on the cold stage\cite{Joshua}.}
\end{figure}

The SPT-3G focal plane consists of ten detector modules, each containing an array of 269 trichroic, dual-polarization pixels on a monolithic 6-inch silicon wafer\cite{Posada} (see Figure~\ref{fig:focal_plane}). Within each pixel, the sky signal couples to a broadband, dual-polarization sinuous antenna. The signal for each polarization is then divided into three frequency bands centered at 95, 150, and 220 GHz by in-line lumped element filters and transmitted via superconducting microstrip to Ti/Au transition-edge sensor (TES) bolometers, for a total of 16,140 detectors in the focal plane.

SPT-3G achieved first light in January 2017. During its first year of operation, work continued to identify possible ways to optimize performance.  In the austral summer season of 2017-2018, several improvements were made, including installing cold lenses with new anti-reflection coating, modifying the readout electronics, and replacing the SQUID transimpedance amplifiers with new low input-impedance versions. Significantly, all ten wafers in the focal plane were replaced with new detector arrays tuned to increase mapping speed and improve overall performance. In this paper we discuss the detector properties and on-sky characterization of the second-year SPT-3G focal plane.

\section{DETECTOR DARK PROPERTIES}

\subsection{TES Normal Resistance}
The normal resistance of the TES, $R_N$, is chosen to be sufficiently larger than any stray impedance in the bias circuit, such that the bolometer can be reliably voltage-biased, while remaining as low as feasible to minimize noise contributions to the readout. For a fixed detector saturation power, a smaller TES resistance requires a smaller voltage bias, which reduces current noise in the readout and decreases demand on the electronics producing the bias signals. In the frequency-multiplexing (fMux) readout scheme employed by SPT-3G, in which each TES is placed in series with an inductor and capacitor to form a resonant circuit, there is an additional motivation to keep $R_N$ low, as excessive resistance increases the bandwidth per channel and can lead to bias-leakage crosstalk between frequency neighbors.  With stray impedances measured to be $\lesssim 0.3~\Omega$\cite{Bender16}, we have set target $R_N \sim 2.0~\Omega$.

\begin{figure} [ht]
\begin{center}
\includegraphics[height=5cm]{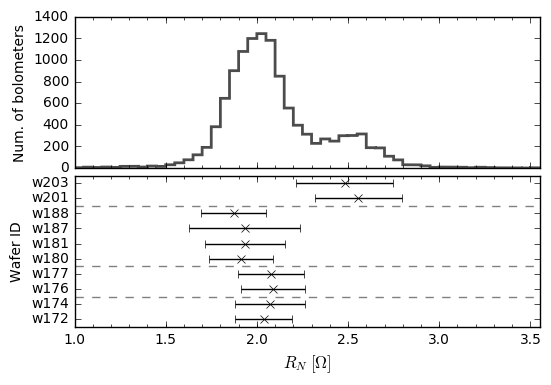}
\end{center}
\caption[rn] 
%>>>> use \label inside caption to get Fig. number with \ref{}
{\label{fig:rn} 
(Top) TES normal resistance distribution on the SPT-3G focal plane, after correcting measurement for effects of fMux readout. (Bottom) Median and standard deviation of $R_N$ per detector wafer. Dashed lines mark changes to the TES layer stack.}
\end{figure}

$R_N$ is primarily determined by the geometry of the 4-layer Ti-Au stack that make up the TES \cite{Faustin}, though it can be affected by other steps in the fabrication process.  The detector wafers for SPT-3G were made across several batches, in which we slightly varied the thicknesses of the TES layers, such that variation in $R_N$ across the focal plane is expected.  Measurements of $R_N$ with the fMux system are also subject to frequency-dependent transfer functions that must be corrected for (see Appendix~\ref{sec:rn_correction}).  The distribution of $R_N$ values is shown in Figure~\ref{fig:rn}; while the last two of the wafers to be fabricated have a slightly higher resistance due to known issues with fabrication at that time, there is sufficient uniformity and proximity to 2~$\Omega$ across the focal plane for our purposes.

\subsection{TES Transition Temperature}
\label{sec: TC}
\begin{figure} [ht]
\begin{center}
\includegraphics[height=5cm]{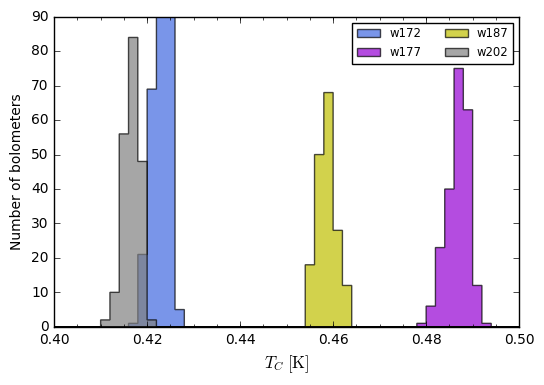}
\end{center}
\caption[tc] 
%>>>> use \label inside caption to get Fig. number with \ref{}
{\label{fig:tc} 
Histograms of TES transition temperature measured on four detector wafers in laboratory test cryostats. These values are representative of the separate batches in which the arrays in the focal plane were fabricated.}
\end{figure}

The desired superconducting transition temperature, $T_C$, of the TES bolometers has a hard lower bound set by the base temperature of the cold stage, which in SPT-3G is approximately 270~mK (for more on the cryogenic design of SPT-3G see Ref~\citenum{Joshua}). $T_C$ should be high enough above the cold stage temperature such that TES performance is unaffected by small temperature fluctuations of the bath, but low enough that thermal fluctuation noise is subdominant to photon noise (see Figure 1 of Ref~\citenum{Faustin}). A critical temperature 1.5-2 times higher than the thermal bath is sufficient, with the exact value depending on the desired saturation power of the bolometers (see section \ref{sec:Psat}). In order to obtain lower saturation powers for the second-year focal plane, target $T_C$ was lowered from 550~mK to $\sim$ 420~mK - 480~mK, varying for successive batches of wafers according to the results of laboratory testing and achieved by altering the thicknesses of the Ti and Au layers in the TES stack. Representative $T_C$ measurements for each of these wafer groups from pre-deployment testing are shown in Figure~\ref{fig:tc}, and show good agreement with their target values.

%The detector wafers are mounted in the standard detector module hardware and blanked off with a metal cover so as to minimize the thermal loading on the detectors.  Measurements of $T_C$ are obtained in two ways: in the first method, a small bias voltage is applied to the TESs, and their resistance measured as the cold stage slowly sweeps in temperature. To account for long thermalization times between the detector array and the cold stage, temperature sweeps are done in both directions, and the average transition temperature between heating and cooling runs is used. In the second method, the saturation power of the TESs is recorded as a function of cold stage temperature, and the result is used to fit for both the detectors' thermal conductance to the bath as well as $T_C$.  Both methods yield very similar results.

\subsection{Saturation Power}
\label{sec:Psat}
\begin{figure} [ht]
\begin{center}
\includegraphics[height=6cm]{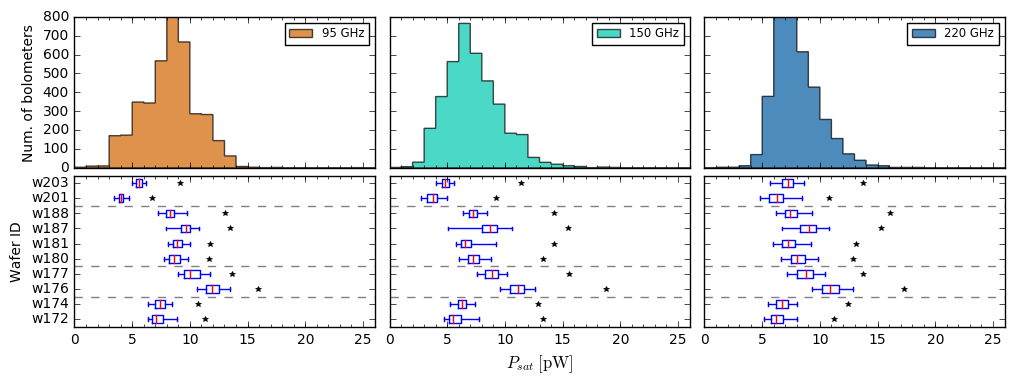}
\end{center}
\caption[psat] 
%>>>> use \label inside caption to get Fig. number with \ref{}
{\label{fig:psat} 
Detector saturation powers measured with incident optical power on-telescope. (Top row) $P_{sat}$ for each observing band across the whole focal plane. (Bottom row) Box plots of $P_{sat}$ for each of the ten detector wafers, with the median, central 50\%, and central 90\% of the distributions indicated. Each wafer has several bolometers that were intentionally left uncoupled to their pixel's antenna; the median value of these dark bolometers' $P_{sat}$ are plotted as black stars. The difference between the optical and dark bolometers is a measure of the optical loading on the detectors.}
\end{figure}

The saturation power, $P_{sat}$, of a TES bolometer is the amount of applied power that will drive the TES out of its superconducting transition and into its normal resistive regime. Our target dark-detector saturation powers are twice the maximum expected optical loading, which in addition to dividing the total incident power roughly equally between optical power and electrical bias power, leaves a reasonable margin of error to avoid saturating the detectors without degrading noise performance or unnecessarily increasing the dynamic range requirements of the electronics. With reasonable assumptions about the detectors' performance we expect $\sim$ (5, 7.5, 10)~pW of optical loading at (95, 150, 220)~GHz\cite{Joshua}, yielding target unloaded $P_{sat}$ values of $\sim$ (10, 15, 20)~pW, respectively. The $P_{sat}$s of the 2017 focal plane were generally higher than this\cite{Anderson}, in part to allow for excess loading should it be higher than expected in new instrument. The first year of operation showed our optical loading to be approximately as-calculated, so we could benefit from lower $P_{sat}$ detectors. As the bolometers' thermal conductance was already maximally low given constraints on the dimensions of the legs supporting the bolometer islands, we lowered $P_{sat}$ for the second year focal plane by decreasing $T_C$, as detailed above.

Measurements of $P_{sat}$ taken on-telescope are shown in Figure~\ref{fig:psat}. Compared to analogous data from the previous year, they show a decrease in median focal plane $P_{sat}$ of (11, 14, 14)~pW at (95, 150, 220)~GHz. While these data were taken with optical loading on the detectors, each wafer includes several bolometers intentionally left uncoupled to their pixel's antenna, and the medians of these dark bolometer $P_{sat}$s are marked with stars in the bottom row of Figure~\ref{fig:psat}.  The average dark $P_{sat}$ values for the three detector bands are (12, 14, 14)~pW.

%, subject to corrections discussed in Appendix~\ref{sec:psat_correction}
\section{OPTICAL PROPERTIES}
\subsection{Time Constants}
The bolometers need to respond to incident radiation fast enough to resolve arcminute-size features at the desired scan speed of the telescope, while being slow enough to be stably operated under negative electrothermal feedback. The former restricts the bolometers' time constants in operation to be smaller than 16~ms, while the latter requires them to be larger than $\sim$~0.50~ms\cite{Irwin98}.  The speed at which the bolometers can absorb heat and then re-equilibrate is set by the heat capacity on the bolometer island and the thermal conductivity of the legs supporting the island.  When the TES is lowered into its transition and operated under negative feedback, this natural time constant $\tau_0$ is sped up by the loopgain of the electrothermal feedback\cite{Irwin98}.  Thus the effective time constant $\tau_{eff}$ of the detectors in operation depends not only on properties of the TES and its thermal connection to the bath, but also on the steepness of the superconducting transition at the point at which the TESs are operated.

\begin{figure} [ht]
\begin{center}
\includegraphics[height=6cm]{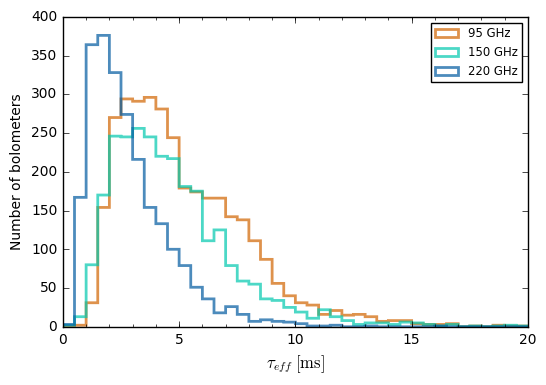}
\end{center}
\caption[bandpasses] 
%>>>> use \label inside caption to get Fig. number with \ref{}
{\label{fig:timeconstants} 
Bolometer optical time constant distribution with detectors tuned to 0.7$R_N$ in their transition.}% Right: The fractional change in median $\tau_{eff}$ with TES operating depth.}
\end{figure}

The bolometers' optical time constants are measured in situ using a chopped thermal source located behind an aperture in the telescope secondary mirror\cite{Zhaodi}. The amplitude of the bolometer response as a function of chopping frequency is then fit to a single-pole low-pass filter model, and the measurement is repeated with the detectors lowered to different depths in their transitions, specified as a fraction of the TESs' normal resistance. We see a $\sim$30\% decrease in median focal plane $\tau_{eff}$ going from 0.9$R_N$ to 0.8$R_N$ with a further $\sim$10\% decrease at 0.7$R_N$, after which $\tau_{eff}$ levels off. There is wafer-to-wafer variation that causes the distribution of  $\tau_{eff}$  values to be somewhat broad, as seen in Figure~\ref{fig:timeconstants}, but all values are within the acceptable range.  For the first half of 2018 we tuned all detectors to 0.7$R_N$, but as each TES can have its tuning parameters adjusted individually, there is room for yet further optimization.  

\subsection{Linearity}
The SPT-3G survey field is a 1500 square-degree patch extending 30$^\circ$ in declination and $6^\mathrm{h}40^\mathrm{m}$ in right ascension. As the telescope moves in pointing elevation, the change in airmass alters the amount of optical loading from the atmosphere on the detectors, shifting the point in their transitions at which the TESs operate. If detector response is not sufficiently linear, this will cause systematic differences in their response between the top of the field and the bottom. To measure the linearity of the detectors over this range, we step the telescope through a series of elevations while illuminating the focal plane with a chopped thermal source. As in Ref~\citenum{George}, we quantify linearity as the maximum percent change in bolometer response over the elevation range, measured across several days in order to reduce the effect of atmospheric variability. Over the full 30 degrees in elevation, we find the median variation across each band to be 2.7\%, 4.3\%, and 1.2\% at 95, 150, and 220~GHz, respectively. 

\begin{figure} [ht]
\begin{center}
\includegraphics[height=6cm]{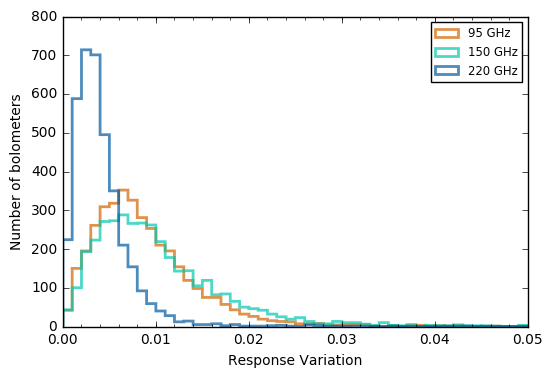}
\end{center}
\caption[linearity] 
%>>>> use \label inside caption to get Fig. number with \ref{}
{\label{fig:linearity} 
The fractional change in bolometer response to a chopped thermal source as the telescope moves through 8 degrees of elevation, slightly larger than the height of each SPT-3G subfield. Bolometers are initially tuned to 0.7$R_N$.}
\end{figure}

While these values show the detectors to be reasonably linear across the full range, the prospect of more uniform performance and other considerations, including uniform depth across a non-rectangular field and changes in telescope focus with elevation, drive us to divide the field up into four equal-height subfields.
Figure~\ref{fig:linearity} shows the stability of the detectors' response from 62 to 70 degrees in elevation, approximately  corresponding to the SPT-3G subfield closest to zenith. The linearity is excellent, with a median variation across each band of 0.77\%, 0.87\%, and 0.35\% at 95, 150, and 220~GHz. Performance degrades slightly as the telescope points closer to the horizon and the TESs are raised to the less-linear portion of their transition. To mitigate this, we re-tune the detectors to 0.7$R_N$ midway through the 1500 square-degree observation, such that linearity is approximately uniform across the full field.

\subsection{Bandpasses}
The detector bandpasses were chosen to fit within the atmospheric transmission windows centered near 95, 150, and 220~GHz, and are achieved via three-pole quasi-lumped-element filters connected to the broadband antenna within the pixels on the detector wafers. We measured the detector bandpasses at the South Pole with a compact Fourier Transform Spectrometer (FTS) mounted on linear actuators and coupled to the optics cryostat window via a mylar beam splitter\cite{Zhaodi}. The output beam of the FTS is designed to match that of a single pixel; however, we found we could get a high signal-to-noise measurement in surrounding detectors with negligible difference in spectral response. This decreased the number of FTS pointings required to obtain a good sampling of detectors across the focal plane. Each wafer has an average of 365 measured detector spectra. 

\begin{figure} [ht]
\begin{center}
\includegraphics[height=6cm]{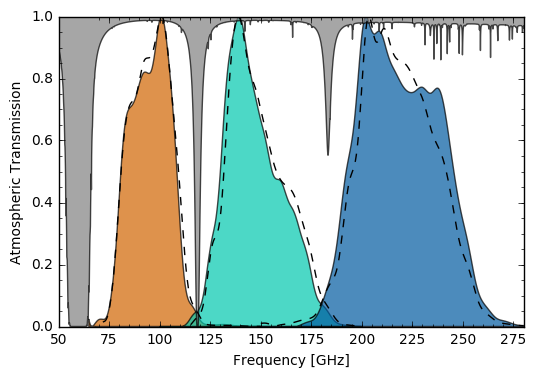}
\end{center}
\caption[bandpasses] 
%>>>> use \label inside caption to get Fig. number with \ref{}
{\label{fig:bandpasses} 
Array-averaged bandpasses for the 2018 (solid, filled) and 2017 (dashed) focal planes, along with the atmospheric transmission model for 0.25~mm PWV. Bands are normalized to one.}
\end{figure}

The raw spectra need to be corrected for frequency-dependent factors in the FTS measurement, namely the $\nu^2$ emission spectrum of the blackbody source used as an input to the FTS and the $\nu^2$ reflection of the mylar beam splitter.  These corrected spectra are plotted in Figure~\ref{fig:bandpasses} and summarized in Table~\ref{tab:bandpass_params}; in general the bands match designed specifications and show excellent consistency with those measured in the previous year. Additional frequency-dependent transmission factors internal to the telescope cause the measured bandpass shape to differ from simulation, including the pixel beams' throughput $A\Omega$, spillover onto the cold Lyot stop, and transmission through the cold optical elements. In particular, the measured difference in the 220~GHz bandpass between 2017 and 2018 is likely due to a change in the anti-reflection coating used on the cold lenses; see elsewhere in these proceedings for a more thorough discussion.\cite{Joshua}$^,$\cite{Andrew} 
 
\begin{table}[ht]
\begin{center}       
\begin{tabular}{|c|c|c|c|c|c|} 
\hline
\rule[-1ex]{0pt}{3.5ex}  \thead{Detector\\Band} &\thead{Number\\Measured} &\thead{Band Center} & \thead{Low Edge} & \thead{High Edge} & $\Delta\nu$ \\
\hline
\rule[-1ex]{0pt}{3.5ex}  95 GHz & 966 & 95.0 $\pm$ 1.3 & 77.0 $\pm$ 3.5 & 110.5 $\pm$ 1.3  & 24.8 $\pm$ 2.6\\
\hline
\rule[-1ex]{0pt}{3.5ex}  150 GHz & 1277 & 147.0 $\pm$ 2.8 & 124.8 $\pm$ 3.6 & 171.1 $\pm$ 4.5  & 31.3 $\pm$ 3.8 \\
\hline
\rule[-1ex]{0pt}{3.5ex}  220 GHz & 1403 & 218.8 $\pm$ 3.3 & 186.1 $\pm$ 8.0 & 252.5 $\pm$ 3.9  & 46.9 $\pm$ 4.7 \\
\hline
\end{tabular}
\caption[bandpass_params]
{\label{tab:bandpass_params}
Array-averaged bandpass parameters, in GHz. The band edges are defined at 25\% of the peak response.}
 
\end{center}
\end{table}

\subsection{Optical Efficiency}
Measurements of detector optical efficiency primarily come from pre-deployment testing in laboratory cryostats. A temperature-controlled blackbody radiation source is installed in the cryostat directly in front of the detector array, and is stepped through a temperature range of approximately 5~K - 20~K. The expected change in optical power is then compared with the change in electrical bias power needed to bias the detectors at a fixed depth in their transition. Detector efficiencies as measured on one wafer are consistent with wafers on the previous focal plane\cite{Zhaodi} and are approximately uniform across the three bands at 0.84 $\pm$ 0.06.
Many factors must be carefully accounted for to obtain an absolute measurement of optical efficiency, including frequency-dependence in the cold electronics transfer function and the transmission of low-pass filters between the thermal source and the detectors. Since all detectors are ultimately calibrated against a known thermal source on the telescope, laboratory measurements of optical efficiency are more useful for detecting large differences within a wafer or between wafers due to errors in fabrication. These relative measurements show all fielded wafers to have similar performance.

Optical efficiency measurements taken on the telescope, done via observations of an astronomical source of known flux, include the effects of transmission through all the elements in the optical chain, such as the cold alumina lenses and their AR-coatings, in addition to the detectors. These end-to-end measurements are discussed in more detail elsewhere in these proceedings\cite{Joshua}, but their results are generally consistent with the efficiency numbers presented here.

\subsection{Polarization Response}
\begin{figure} [ht]
\begin{center}
\includegraphics[height=5cm]{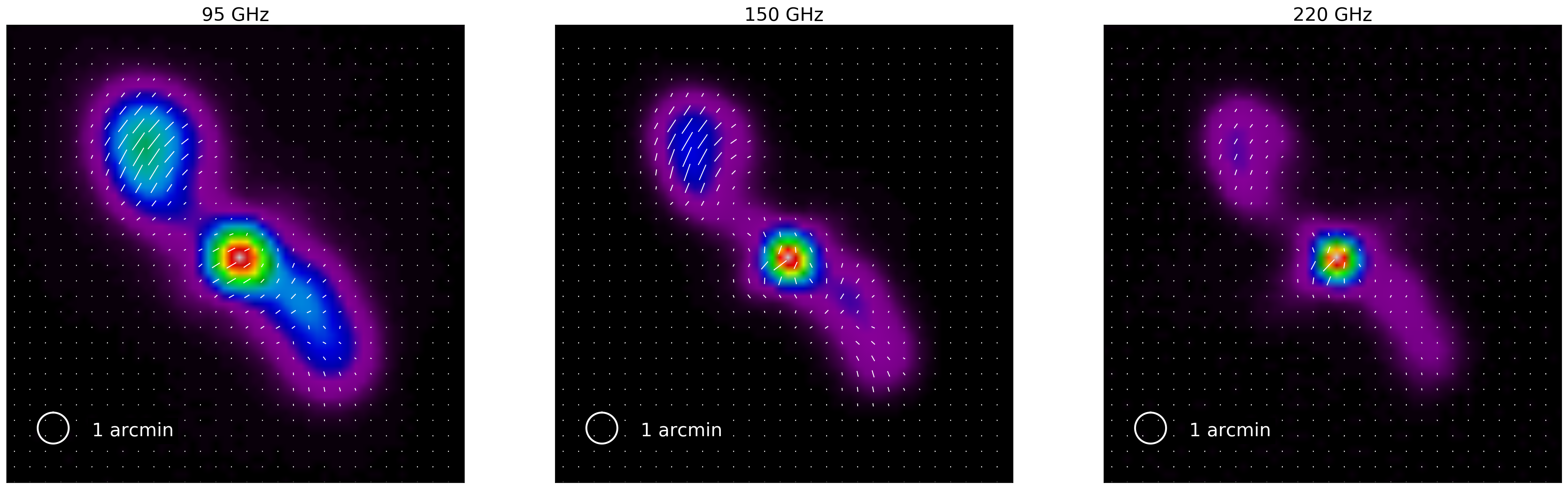}
\end{center}
\caption[cenA] 
%>>>> use \label inside caption to get Fig. number with \ref{}
{\label{fig:cenA} 
Preliminary three-band map of CenA from early 2018 SPT-3G data showing polarization (vectors) and intensity (color).}
\end{figure} 

Characterization in laboratory cryostats has shown the polarization efficiency of SPT-3G pixels to generally be $>$90\% \cite{Zhaodi}, but per-detector polarization angles still need to be measured on-telescope. For the SPTpol camera, polarization calibration was done primarily via a polarized source positioned 3~km away from the telescope\cite{George}.  For SPT-3G, we plan to obtain polarization angle measurements via observations of a polarized astronomical source, Centaurus A (Figure~\ref{fig:cenA}). Cen A is a radio galaxy with a bright compact nucleus and collimated jets that expand into large radio lobes, the polarization properties of which have been previously measured at roughly the angular scale and frequencies of interest from the South Pole by the QUaD experiment\cite{QUaD}, finding one of the radio lobes to have a polarization fraction of about 20\%. Final polarization calibration will include cross-correlating with other CMB experiments, including SPTpol and \emph{Planck}.

\section{Yield}

\begin{figure} [ht]
\begin{center}
\includegraphics[height=6cm]{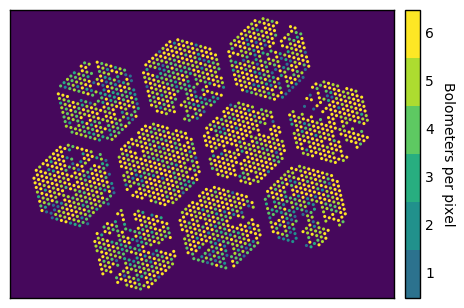}
\end{center}
\caption[ptsrc_fp] 
%>>>> use \label inside caption to get Fig. number with \ref{}
{\label{fig:ptsrc_fp}
Distribution of active bolometers across the fielded 2018 SPT-3G focal plane.}
\end{figure}

There are 16,140 bolometers in the SPT-3G focal plane, though the design of the detector wafers and other cold components limit the number of wired bolometers to 15,720. For the 2018 focal plane 10\% of these were either open or shorted to another detector at room temperature. While we choose to not attempt to operate these bolometers, if the $\sim$M$\Omega$ resistance of an open detector becomes superconducting once cryogenic or the $\sim$k$\Omega$ short between detectors does \textit{not} become superconducting once cryogenic, those bolometers are in principle recoverable. Bolometers channels that do not show a resonance near the expected frequency in a network analysis (see e.g. Ref~\citenum{Bender16}) due to a break in the cold circuit chain cause another reduction in yield; see Table~\ref{tab:yield}. Small gains can be made via manual verification of the network analysis results. To quantify on-sky bolometer yield we take the number of detectors that pass a signal-to-noise cut on their response to a chopped calibration source.  This eliminates dark bolometers (2\% of identified resonances) as well as TESs that do not show a transition or are otherwise abnormal and thus dropped from array tuning. Approximately 11,400 bolometers have passed this cut to date, with 8\% fewer passing in a typical observation. The distribution of live bolometers is shown in Figure \ref{fig:ptsrc_fp}.

\begin{table}[ht]
\begin{center}       
\begin{tabular}{|c|c|c|c|c|} 
\hline
\rule[-1ex]{0pt}{3.5ex} \thead{Wired\\bolometers} &\thead{Passed room temp.\\connectivity} & \thead{Identified\\resonance} & \thead{Biased} &\thead{Optically\\responsive} \\
\hline
\rule[-1ex]{0pt}{3.5ex} 15,720 & 90\% & 85\% & 76\% & 72\% \\
%\rule[-1ex]{0pt}{3.5ex} 15,720 & 14,195 & $\sim$13,431 & $\sim$11,388 \\
\hline
\end{tabular}
\caption[yield]
{\label{tab:yield}
Bolometer yield for SPT-3G 2018 focal plane. "Biased" denotes detectors that we attempt to operate, which excludes e.g. TESs that do not show a transition or that have a resonant peak overlapping another detector. "Optically responsive" includes bolometers that have passed a signal-to-noise cut on their response to a chopped calibration source.}
\end{center}
\end{table}

\section{CONCLUSION}

The second year of SPT-3G operation includes several improvements to the instrument, including a brand new focal plane with detector wafers better tuned for optimal performance. Improved fabrication techniques for the Ti-Au TES have allowed the transition temperature and saturation powers of the bolometers to be fine-adjusted to approach our target values. Other parameters of the detectors, such as TES normal resistance, bolometer time constants, filter bandpasses, and detector optical efficiency, were already within acceptable ranges and remain so in the new detector wafers. SPT-3G began a five-year survey of a 1500 square-degree field in March 2018.

\appendix
\section{Correcting Resistance Measurements}
\label{sec:rn_correction}
Dedicated four-wire measurement of TES resistance for SPT-3G can only be conducted on a small number of test structures or fragments of detector arrays broken for this express purpose\cite{Faustin}, whereas $R_N$ measurements utilizing the fMux system are made on every TES in the focal plane as part of regular operation. However, the latter is subject to impedances in the cold wiring and frequency-dependent transfer functions that do not affect the former; four-wire measurements of TESs from a range of wafer radii show $R_N$ to be uniform within $0.1~\Omega$, while values obtained during regular operation exhibit significantly more spread. 

\begin{figure} [ht]
\begin{center}
\includegraphics[height=4cm]{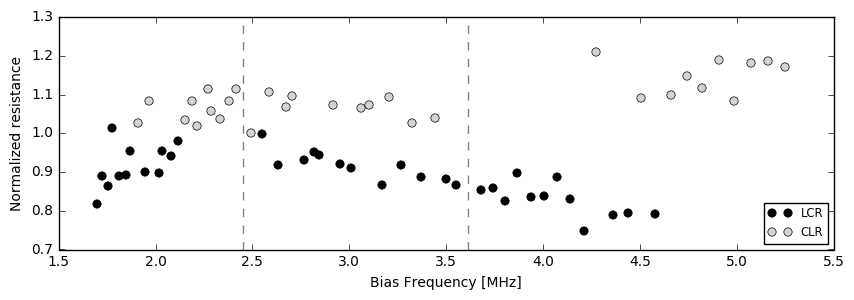}
\end{center}
\caption[highR_ch] 
%>>>> use \label inside caption to get Fig. number with \ref{}
{\label{fig:highR_ch} 
The relation between bolometer channel frequency and apparent resistance. Black points represent channels with resonant circuit elements ordered inductor-capacitor-TES, while gray points represent channels ordered capacitor-inductor-TES. Overall normalization is set by two 1~$\Omega$ resistors on the LC boards. The dashed vertical lines mark the separation of the 95, 150, and 220~GHz detectors into the lower, middle, and upper regions of the bias frequency range, respectively.}
\end{figure}

One factor contributing to this increased variation is the effect of alternating the order of the lithographed inductor and capacitor on the monolithic resonant filter ($LC$) chip.  The checkerboard pattern increases the spatial separation between inductors to reduce crosstalk via mutual-inductance\cite{Hattori}, but the $LCR$ and $CLR$ orderings also create two distinct frequency-resistance transfer functions.  This general behavior can be reproduced in circuit simulations by modeling each distributed inductor and capacitor element as having a $\sim$1~pF capacitive coupling to the ground plane of the $LC$ chip, which in turn is coupled to cryostat ground.  The average of resistance measurements from all $LC$ chips was used to empirically construct a bolometer channel-resistance relation, shown in Figure~\ref{fig:highR_ch}, that can be utilized to give more accurate measures of TES normal resistance with this implementation of the fMux system.

\begin{figure} [ht]
\begin{center}
\includegraphics[height=5cm]{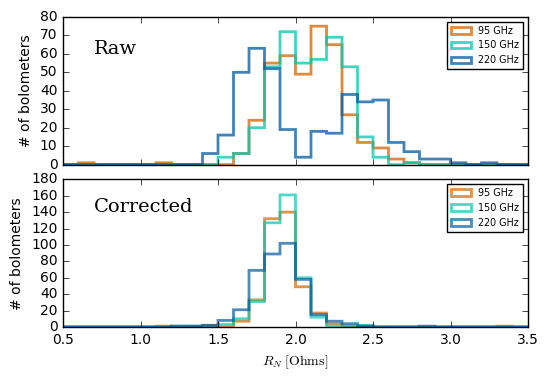}
\end{center}
\caption[rn_correction] 
%>>>> use \label inside caption to get Fig. number with \ref{}
{\label{fig:rn_correction} 
Normal resistance measured on one detector wafer before correcting for the $LCR-CLR$ transfer function (top) and after the correction is applied (bottom). Post-correction the distribution is less broad and no longer bimodal within each detector band.}
\end{figure}

Figure~\ref{fig:rn_correction} shows measurements of $R_N$ for one of the SPT-3G detector wafers. The upper panel shows uncorrected values in which the spread in $R_N$ is apparent, as is the pronounced bimodality in the 220~GHz bolometers. The lower panel shows the same data after correcting for the frequency-resistance relation; though some residual scatter remains, the distribution of $R_N$ is much tighter, no longer bimodal, and doesn't display any notable difference between detector bands. The absolute calibration of this correction is set by two 1~$\Omega$ resistors on the LC boards, and after correcting for the measured $\sim0.3~\Omega$ stray resistance yields $R_N$ distributions approximately matching the few number of 4-wire measurements that were performed. More precise numerical results based on circuit simulations is a subject for further study, but for future chips it is recommended to shift the rows of filter elements with respect to each other instead of alternating the order of the circuit elements, as this will achieve the same crosstalk mitigation without the additional transfer function.

\acknowledgments % equivalent to \section*{ACKNOWLEDGMENTS}       

The South Pole Telescope program is supported by the National Science Foundation (NSF)  through grant PLR-1248097.
Partial support is also provided by the NSF Physics Frontier Center grant PHY-1125897 to the Kavli Institute of Cosmological Physics at the University of Chicago, the Kavli Foundation, and the Gordon and Betty Moore Foundation through grant GBMF\#947 to the University of Chicago.
Work at Argonne National Lab is supported by UChicago Argonne LLC, Operator of Argonne National Laboratory (Argonne). Argonne, a U.S. Department of Energy Office of Science Laboratory, is operated under contract no. DE-AC02-06CH11357. We acknowledge R. Divan, L. Stan, C.S. Miller, and V. Kutepova for supporting our work in the Argonne Center for Nanoscale Materials.
Work at Fermi National Accelerator Laboratory, a DOE-OS, HEP User Facility managed by the Fermi Research Alliance, LLC, was supported under Contract No. DE-AC02-07CH11359. 
NWH acknowledges support from NSF CAREER grant AST-0956135. 
The McGill authors acknowledge funding from the Natural Sciences and Engineering Research Council of Canada, Canadian Institute for Advanced Research, and the Fonds de recherche du Qu\'ebec Nature et technologies.
Vieira acknowledges support from the Sloan Foundation.

% References
%\bibliography{report} % bibliography data in report.bib
%\bibliographystyle{spiebib} % makes bibtex use spiebib.bst

\end{document}